\documentstyle[aps,epsfig,pre,twocolumn]{revtex}
\begin{document}
\draft

\title{Interface dynamics from experimental data}
\author{Achille Giacometti$^{(1)}$, and Maurice Rossi$^{(2)}$}

\vskip 2.0cm
\address{$^{(1)}$
INFM Unit\'a di Venezia, Dipartimento
di Scienze Ambientali, Universit\'a di Venezia,\\
Calle Larga Santa Marta DD 2137, I-30123 Venezia-Italy}
\address{$^{(2)}$ Laboratoire de Mod\'elisation en M\'ecanique,
Universit\'e de Paris VI, 4 Place Jussieu, F-75252 Paris Cedex 05, France}

\date{\today}

\maketitle


\begin{abstract}
A novel algorithm is envisaged to extract the coupling parameters
of the Kardar-Parisi-Zhang (KPZ) equation from experimental
data. The method hinges on the Fokker-Planck equation combined
with a classical least-square error procedure. It takes properly into account
the fluctuations of surface  height  through a deterministic equation for
space correlations. We apply it to the 1+1 KPZ
equation and carefully  compare its results  with those obtained by previous
investigations. Unlike previous approaches, our method does not require large
sizes   and  is stable under a modification  of
sampling time of observations.  Shortcomings associated to standard
discretizations of the continuous KPZ  equation are also  pointed out  and
possible future perspectives are finally analyzed.
\end{abstract}
\pacs{64.60.Ht,05.40.+j,05.70.Ln}
%

\section{Introduction}
\label{sec:introduction}

Inverse  techniques have a wide range of applicability ranging from
geophysics to nonlinear time analysis and statistics
\cite{Weigend94}.  The common philosophy behind these methods  is the
extraction of  equations  of motion  starting from successive experimental
time series  of some  dynamical variable in addition to
basic assumptions such  as determinism.
If a reasonably general form of the equations is guessed either by
symmetry arguments or by general considerations, the ``true'' parameters are
then determined by minimizing a cost function  quantifying  the distance
between
experimental observations and corresponding reconstructed  quantities, the
latter being implicitly dependent upon the parameters.  Among such approaches,
the Least Square  method  is the most popular one.

A typical system which can be treated using reconstruction techniques is
the case where an {\it observational noise} is superimposed to a standard
{\it deterministic} evolution.  In this case the system is expected to evolve
under the action of  deterministic system and stochasticity come only from our
measurement apparatus. The particular case where the dynamics underlying the
system is  chaotic has also received considerable attention
due to its widespread
occurence in natural systems \cite{Abarbanel93}, and the importance of treating
the presence of the noise with due care has already been emphasized
\cite{Kostelich93}.

The alternative possibility of {\it dynamical noise} occurs  whenever the
noise is a built-in component of the equations of motion. This is a far more
difficult problem since one has to deal with stochastic rather than
deterministic equations.
An important such case, which is rather ubiquitous in nature, is the
Langevin dynamics where  variables evolve  subject  both to  dissipative
generalized forces and to a fluctuating part \cite{Gardiner90}. In this latter
instance,  the presence of  dynamical  noise can drastically modify the
dynamics and hence hampers the efficiency of the usual
reconstruction techniques
based on deterministic ideas\cite{Battiston99}.

In our work, we  focus on a particular class of Langevin dynamics
which has its origin in a seminal paper on interface dynamics \cite{Kardar86}
but has ever since displayed relations to a variety of physical systems
such as for instance bacterial colonial growth, immiscible fluids,
directed polymer and superconductors \cite{Krug97}.

The Kardar-Parisi-Zhang (KPZ) equation \cite{Kardar86} was introduced
as a coarse-grained mesoscopic description for the growth of a rough surface
under the deposition of particles driven by gravity.
The crucial ingredient introduced in the KPZ and not present in the
corresponding linear counterpart, namely the Edward-Wilkinson
(EW) equation \cite{Edward82}, is a non-linear term which
takes into account the fact that the growth is normal to the surface.
The KPZ equation can be mapped into various other models. A Cole-Hopf
change of variables maps it into a directed polymer diffusion equation
subject to a random potential \cite{Huse85}, while the identification of
the local
gradient with a velocity leads to the Burgers's equation for a
vorticity-free velocity field \cite{Forster77}. Furthermore it is believed that
the KPZ equation has the same large-scale behavior of the
Kuramoto-Sivashinsky equation in $1+1$ dimensions \cite{Yakhot81}, while in
higher
dimensionality the situation is much less clear \cite{Bogosian99}.
Nonetheless, in spite of the gigantic effort devoted to the KPZ equation
in the past decade, a complete understanding of its properties is still
lacking.

The aim of the present paper is to introduce a new inverse approach to the KPZ
equation. A previous attempt due to Lam and Sander \cite{Lam93} was  based
on the standard Least Square (LS) reconstruction method. These authors
used this approach directly on numerically simulated experimental surfaces
without a preventive test of the performance of the method itself.
Lam and Shin \cite{Lam98_1}
subsequently showed that the standard discretizations used in \cite{Lam93} was
not adequate. We shall argue below that, even with the improvements given in
\cite{Lam98_1}, the classical identification procedure devised in
Ref.\cite{Lam93}
is not properly suited
for Langevin dynamics since it is  based on deterministic equation ideas.
By an explicit computation using the LS technique applied
to a $1+1$ KPZ equation, we shall review their method and point
out what we consider its main deficiencies.

We then go on to introduce a different approach based on
the Fokker-Plank equation (FPE) associated to each Langevin equation
\cite{Risken89},\cite{Gardiner90}. The
advantage of this viewpoint is that one can construct {\it deterministic}
relations among correlation functions which however still carry informations
regarding the fluctuating nature of the original quantities. Those equations can
then be easily analyzed within a least squares framework as in the LS method.

The paper is organized as follows.
In Sec.~\ref{sec:interface}, the  KPZ equation  is rapidly recalled along
with its numerical real space approximations in $1+1$ dimensions while
the LS approach is reviewed in section~\ref{sec:least_squares}.
Sec.~\ref{sec:stochastic} contains the basic equations of our modified method
which is then applied in Sec.~\ref{sec:identification}.
Numerical results are then given in Sec.~\ref{sec:results} and some concluding
remarks are provided in Sec.~\ref{sec:conclusions}. More technical points
are finally confined in the Appendices. Appendix \ref{sec:appendixa} shows
why the least
square method fails for sufficiently large noise amplitudes and Appendix
\ref{sec:appendixb}  presents some
results concerning  renormalized interfaces and their corresponding
renormalized equations.

\section{Interface dynamics}
\label{sec:interface}

We consider a one-dimensional line of total length $L$ and a surface
of height $h(x,t)$ at position $x$ and time $t$. The continuum
$1+1$ KPZ equation then reads:

\begin{eqnarray}
 \partial_t  h(x,t) & = & c  + \nu \partial_x^2 h(x,t)
+\frac{\lambda }{2} \bigl[ \partial_x h(x,t) \bigr]^2 + \eta (x,t),
\label{interface1}
\end{eqnarray}

where  $ \eta (x,t) $ is an uncorrelated white noise
\begin{eqnarray}
\left \langle \eta (x,t) \eta (x',t') \right \rangle &=&
2 D \delta(x-x')~\delta(t-t').
\label{interface2}
\end{eqnarray}
The average  $\left \langle \;\;\;  \right \rangle$ is taken on different
realizations of the noise. In Eq.(\ref{interface1}) and (\ref{interface2}),
$c$, $\nu$, $\lambda$ and $D$ are coupling parameters ($c$ is often set to zero
because of the invariance of Eq.(\ref{interface1}) under rescaling $h \to
h+ct$).
For $\lambda=0$,  Eq.(\ref{interface1}) reduces to the Edward-Wilinson
(EW) equation \cite{Edward82} which can be solved exactly.

In writing Eq.(\ref{interface1}) either a
regularization in the correlation given in Eq. (\ref{interface2})
(such as for instance
a spatially correlated noise ) or the introduction of a minimal length
scale $a$ is always tacitly assumed.
In the latter case, one is then naturally led to consider a discretization
of the continuum equation at a given cutoff length scale $a$.  In that   case,
(a) the noise term $ \eta (x,t)$ is discretized
\begin{eqnarray}
\eta_i(t)&=& \sqrt{\frac{D}{a}} \theta_i(t),
\label{interface3}
\end{eqnarray}
where $\theta_i(t)$ is  a  random noise
\begin{eqnarray}
\left \langle \theta_i (t) \theta_j (t') \right \rangle &=&
2 \delta_{i,j}~\delta(t-t').
\label{interface4}
\end{eqnarray}
with $\delta_{i,j}$ the Kronecker symbol; and  (b)  Eq. (\ref{interface1}) is
written for a discrete variable $h_i(t)$ ($i=1,\ldots,N=L/a$) with periodic
boundary conditions
\begin{eqnarray}
\frac{d h_i }{dt}&=& c+ \nu_{\mathrm{eff}}~F_i^{\nu}[h] +
\frac{\lambda_{\mathrm{eff}}}{2}~F_i^{\lambda}[h] +\sqrt{D_{\mathrm{eff}}}
\theta_i(t).
\label{interface5}
\end{eqnarray}
Here $\nu_{\mathrm{eff}}=\nu/{a^2}$, $\lambda_{\mathrm{eff}}=\lambda/{a^2}$,
$D_{\mathrm{eff}}=D/a$. $F_i^{\nu}[h]$ and $F_i^{\lambda}[h]$ are proper
discretizations of the linear $\partial_x^2 h$ and non-linear
$(\partial_x h)^2$
terms respectively. We note that the exact meaning of ``proper discretization''
has been the object of some
investigations \cite{Beccaria94,Newman97,Newman96,Lam98_2}.

In all practical applications, a further  temporal discretization
\cite{Risken89}, \cite{Mannella89} is also   performed on
Eq. (\ref{interface5})

\begin{eqnarray}
h_i  (t + \delta t) & =& h_i (t) + \delta t ~\Bigl(c+
\nu_{\mathrm{eff}}~F_i^{\nu}[h(t)] +
\frac{\lambda_{\mathrm{eff}}}{2}~F_i^{\lambda}[h(t)]~\Bigr)  + \sqrt {2
D_{\mathrm{eff}} \delta t}~r_i,
\label{interface6}
\end{eqnarray}
where $r_i $ is a Gaussian random generator of unit variance and
$\delta t$ the discretization time step.

In $d=1+1$ it is known that the steady state solution
$P[h]$ for the probability distribution of the heights in the KPZ
equation  is identical to the EW stationary distribution due to a
fluctuation-dissipation theorem \cite{Forster77}. It was shown
\cite{Lam98_2},\cite{note1} that the correct stationary
discrete probability namely
\begin{eqnarray}
P[h] &=& {\cal N}^{-1} \exp
\biggl[ -\frac{1}{2} \frac{\nu}{D a}
\sum_{i=1}^{N} (h_{i}-h_{i+1})^2 \biggl],
\label{interface7}
\end{eqnarray}
where ${\cal N}^{-1}$ is a normalization factor, can be obtained
by taking
\begin{eqnarray}
F_i^{\nu}[h] &=& h_{i+1}+h_{i-1}-2 h_i,
\label{interface8}
\end{eqnarray}
and
\begin{eqnarray}
F_i^{\lambda}[h] &=& \frac{1}{3} \Bigl[ (h_{i+1}-h_i)^2+
(h_{i+1}- h_i)(h_i-h_{i-1})+(h_i-h_{i-1})^2 \Bigl].
\label{interface9}
\end{eqnarray}
The standard choice $F_i^{\lambda}[h] = 1/4~ (h_{i+1}-h_{i-1})^2$
on the other hand fails to reproduce Eq. (\ref{interface7}) and
suffers of other
problems as well \cite{Newman96}. A necessary (albeit not
sufficient) condition for the  identification  with the continuum counterpart
Eq. (\ref{interface1}), is clearly that the correct steady state
(i.e. independent of
$\lambda$) is recovered. For this reason, we shall exploit for   the new
identification procedure as well as for the LS scheme, equations
(\ref{interface8}) and (\ref{interface9}) hereafter instead of the standard
choice which was used in \cite{Lam93}.

\section{The Least Squares error model method}
\label{sec:least_squares}

Before introducing our method we first review the LS
error method used in Ref.\cite{Lam93}.  We consider
experimental surfaces  coarse-grained  at length scale $a$
described by the interface heights  $h_i^{\mathrm{obs}}(t)$, ($i=1,\ldots,N$)
which are  sampled $M$ times i.e. at discrete   times $t=t_k=k \Delta t$
($k=1,\ldots,M+1$).
Note that the sampling time  $\Delta t$ is the time interval
between two experimental observations and it is clearly different from the
discretization time $\delta t$ of Eq.(\ref{interface6}). For
surfaces obtained by numerical simulations $\Delta t$ is typically 
a multiple of  $\delta t$. We note that in Ref.\cite{Lam93},
the authors used $\Delta t$ equal to $  \delta t$
which is a rather particular case.

For the sake of simplicity,  we assume here that measurements
are free from observational noise.
It must be emphasized that, in the presence of measurement noise,
our method performs a priori better then the LS scheme since
it is based on spatial-averaged values which are less affected
by errors on local height measurements.

Our purpose is to determine the coefficients $c$, $\nu $, $\lambda$, $D$
at the given  length scale $a$ in Eq.(\ref{interface3}).

Let us first neglect the dynamical  noise in Eq.(\ref{interface5}).
We then obtain a standard identification problem of the coupling parameters
governing a deterministic non-linear equation which can be cast in the
compact form:

\begin{eqnarray}
\frac{dh_i }{dt} &=& \sum_{\alpha = 1}^{p} \mu_{\alpha} F^{\alpha}_i [h],
\label{least_squares1}
\end{eqnarray}
where in the present  case $p=3$ and $\mu_1 = c$, $\mu_2= \nu_{\mathrm{eff}}$,
$\mu_3=\lambda_{\mathrm{eff}}$, (\ref{interface8})  and (\ref{interface9}) 
are used for $F^{2}_i [h]$ and $ F^{3}_i [h] $  whereas  $F^{1}_i [h] =1$.

Optimal parameters  are then determined by minimizing  a cost
function ${\cal J}$ such as the sum-square difference

\begin{eqnarray}
{\cal J} &=& \frac{1}{N M} \sum_{k = 1}^{M}
\sum_{i = 1}^{N}  \Bigl[ h_i^{\mathrm{obs}}(t_{k+1}) -
h_i^{\mathrm{pred}} (t_{k+1}) \Bigr]^{2},
\label{least_squares2}
\end{eqnarray}

which quantifies  the distance between  experimental observations
$h_i^{\mathrm{obs}}(t_k)$ and  equivalent reconstructed   quantities
$h_i^{\mathrm{pred}}(t_k)$. The
latter quantities are computed from Eq.
(\ref{least_squares1}) for given
parameters and are thus  generally implicit functions of the   parameters.
However, if  the sampling time $ \Delta t$  is small enough,  then
$F^{\alpha}_i [h] $ are nearly constant  between two
measurements and the amplitudes $h_i^{\mathrm{pred}} (t_{k+1})$
can be related to  the parameters $\mu_{\alpha}$ by

\begin{eqnarray}
h_i^{pred} (t_{k+1}) & =& h_i^{obs}(t_k) + \Delta t
\sum_{\alpha = 1}^{p} \mu_{\alpha} F^{\alpha}_i
[h^{\mathrm{obs}}(t_k)]
\label{least_squares3}
\end{eqnarray}

In this case, the cost function ${\cal J}(\{\mu \})$
itself becomes explicit and
quadratic in the  parameters. Optimal parameters can thus be evaluated
through a simple matrix inversion. Indeed
the extremal value  of ${\cal J}$ is inferred by

\begin{eqnarray}
\frac{\partial {\cal J} }{\partial \mu_{\alpha}}|_{\mu^{*}}& =& 0.
\label{least_squares4}
\end{eqnarray}
The solution for  the optimal parameters $\{\mu^{*}\}$
is then given by a matrix equation:

\begin{eqnarray}
\mu_{\alpha}^{*} &=& \sum_{\beta = 1}^{p} A^{-1}_{\alpha \beta}
\cdot B_{\beta},
\label{least_squares5}
\end{eqnarray}
where we have defined
\begin{eqnarray}
A_{\alpha \beta} &=& \frac{1}{N M} \sum_{k = 1}^{M}
\sum_{i = 1}^{N}   F^{\alpha}_i F^{\beta}_i,
\label{least_squares6}
\end{eqnarray}
\begin{eqnarray}
B_{\alpha}&=& \frac{1}{N M} \sum_{k = 1}^{M}
\sum_{i = 1}^{N} \Bigl[ \frac{h_i^{obs}(t_{k+1}) -
h_i^{obs}(t_k) } {
\Delta t }   \Bigr] F^{\alpha}_i,
\label{least_squares7}
\end{eqnarray}
and where functions $F^{\alpha}_i$ are clearly expressed at
$h_1^{\mathrm{obs}}(t_{k}),\ldots,h_N^{\mathrm{obs}}(t_{k})$.

This classical least squares method is an easy and  natural approach
and it works fairly well in the absence of any noise. In the
presence of noise however, it has its main drawback in the fact that
it  approximates time derivatives by finite differences. If the
dynamics is governed by a deterministic equation and  measurements are
performed with a negligible observational  noise,
this simply imposes the choice of a sampling time much
smaller than the characteristic or relaxation time of the process.

Lam and Sander \cite{Lam93} assumed that if the sampling time $ \Delta t$ is
small
enough, the above method could be extended to a Langevin equation (i.e. with
dynamical noise). The amplitude of the noise can then be inferred from
Eq.(\ref{least_squares2}) when $J$ is taken at the minimum values of the
parameters, that is
\begin{eqnarray}
D&=& \frac{1}{2 \Delta t}  a J(\{\mu^{*}\})
\label{least_squares8})
\end{eqnarray}

However it was already observed in dynamical systems that even with pure
measurement noise the above  method can cause  large errors.
This is expected to be the case
for dynamical noise as well.  Two main reasons for this could be  advocated.
First if $\Delta t$ is too large, the linear approximation
(\ref{least_squares3}) which
explicitly relates the observed quantities breaks down.
Because  of the dynamical
noise term, this happens a priori  for  shorter times intervals in a Langevin
equation compared with its deterministic counterpart.  Second, even in the
favorable case in which   $\Delta t$ is small ,  such a  method is
efficient only if large sizes and small noise amplitudes are used. This is
explained in Appendix \ref{sec:appendixa} where a simple zero-dimensional case
is explicitly worked out with the method of Lam and Sander.

\section{Stochastic approach for Model identification.}
\label{sec:stochastic}
We now turn to our method which is based on the simple observation that
all the information present in the Langevin equation (\ref{interface5})
are also contained in the corresponding Fokker-Planck equation (FPE)
\cite{Risken89}:
\begin{eqnarray}
\partial_t P[h,t] &=&
 \sum_{i=1}^{N} D_{\mathrm{eff}} \frac{\partial^2}{\partial h_i^2} P[h,t]
 - \sum_{i=1}^{N} \frac{\partial}{\partial h_i} \Bigl( F_i[h]~P[h,t] \Bigl),
\label{stochastic1}
\end{eqnarray}
where
\begin{eqnarray}
 F_i [h] &=& c + \nu_{\mathrm{eff}} F_i^{\nu}[h]+
\frac{1}{2} \lambda_{\mathrm{eff}} F_i^{\lambda}[h],
\label{stochastic2}
\end{eqnarray}
and \cite{Goldenfeld93}
\begin{eqnarray}
P[h,t] &=& \left \langle \prod_{i=1}^N \delta(h_i -h_i(t)) \right \rangle,
\label{stochastic3}
\end{eqnarray}
where the solution $h_i(t)$ is associated to a particular noise configuration
$\theta_i(t)$.

In equation  (\ref{stochastic1}) the second term on the r.h.s.
characterizes the deterministic  behavior  of  the
system whereas the  first term contains stochastics effects.
We derive a first general equation involving the parameters
$c$ and $\lambda$.
Using  Eq. (\ref{stochastic1}),  the time derivative of the ensemble average of
 $h_i(t)$  can be easily shown to be:
\begin{eqnarray}
\frac { d \left \langle  h_i (t)  \right \rangle } {dt} &=&
\int ~ {\cal D} h  ~  F_i[h]~  P[h,t],
\label{stochastic4}
\end{eqnarray}
where ${\cal D} h \equiv \prod_{i=1}^N ~ d h_i$.
If we denote by  $g^{(1)}(t) = \frac{1}{N} \sum_i \left \langle h_i(t)
\right \rangle $ the  mean
height at time $t$ averaged over the noise, its time derivative can be
written after some simple algebra as
\begin{eqnarray}
\frac{dg^{(1)}(t)}{dt} &=&  c+ \frac{\lambda_{\mathrm{eff}}}{6}
\bigl[ 2 g_0^{(2)} (t)+g_1^{(2)} (t) \bigr],
\label{stochastic5}
\end{eqnarray}

where we have defined

\begin{eqnarray}
g_l^{2} (t) &=&  \frac{1}{N}  \sum_{ i=1 }^{N} \left \langle \delta h_{i}(t)
\delta h_{i+l}(t) \right \rangle,
\label{stochastic6}
\end{eqnarray}
in the variables  $\delta h_i=  h_{i} -  h_{i+1}$.
Note that there are  only $N-1$ independent $\delta h_i $ variables
due to periodic boundary conditions and to the fact that
$\sum_{ i=1 }^{N }\delta h_i=0$.

The above result prompts a convenient change of variables from $h_1,..,h_N$
to  $ \delta h_1,..,\delta h_{N-1},\bar{h} \equiv 1/N \sum_{i=1}^N h_i$
followed by an integration over $\bar{h}$.
Physically this is related to the fact that our system is infinitely degenerate
with respect to the average height.
Note that the stationary
probability Eq. (\ref{interface7}) is now Gaussian and well defined
in the new variables $\delta h_1,\ldots,\delta h_{N-1}$. The corresponding
probability $\widetilde{P}[\delta h]$ is the solution of a modified
FPE:

\begin{eqnarray}
\label{stochastic7}
\partial_t \widetilde{P}[\delta h,t] &=&
2 D_{\mathrm{eff}}  \sum_{i=1}^{N-1}
\frac{\partial ^2}{\partial \delta h_i^2}
\widetilde{P}[\delta h,t] \\ \nonumber  
&-&
\sum_{i=1}^{N-1}
\frac{\partial}{\partial \delta h_i} \Bigl( G_i[\delta h]
\widetilde{P}[\delta h,t] \Bigr)  -
2 D_{\mathrm{eff}} \sum_{i=2}^{N-1}
\frac{\partial ^2}{\partial \delta h_i \partial \delta h_{i-1} }
\widetilde{P}[\delta h,t],
\end{eqnarray}
where we have defined
\begin{eqnarray}
G_i&=&F_i-F_{i+1}=\nu_{\mathrm{eff}} G_i^{\nu}+
\frac{1}{2} \lambda_{\mathrm{eff}} G_i^{\lambda},
\label{stochastic8}
\end{eqnarray}
with
\begin{eqnarray}
G_i^{\nu}&=& \delta h_{i+1}+\delta h_{i-1}-2 \delta h_i,
\label{stochastic9}
\end{eqnarray}
and
\begin{eqnarray}
G_i^{\lambda}&=& \frac{1}{3} \Bigl[ \delta h_{i-1}^2-
\delta h_{i+1}^2-\delta h_i (\delta h_{i+1}-\delta h_{i-1}) \Bigr].
\label{stochastic10}
\end{eqnarray}
We are now in the position to derive  our second basic result.

Integrating Eq. (\ref{stochastic7})  over all variables but (say)   $ \delta
h_j$, one gets, for the single variable  probability
\begin{eqnarray}
 p(\delta h_j)&=& \int {\cal D} \delta h_j
  \widetilde{P} [\delta h,t],
\label{stochastic11}
\end{eqnarray}
where the shorthand notation ${\cal D} \delta h_j=\prod_{i \ne j=1}^{N-1}
d\delta h_i$
was again exploited, the following equation:
\begin{eqnarray}
\partial_t p(\delta h_j,t)& =&
2 D_{\mathrm{eff}}   \frac{\partial ^2}{\partial \delta h_j^2} p(\delta h_j,t)
 -  \frac{\partial}{\partial \delta h_j} \pi(\delta h_j,t),
\label{stochastic12}
\end{eqnarray}
where the non-local term $\pi(\delta h_j,t)$ is defined as
\begin{eqnarray}
  \pi(\delta h_j,t) &=& \int {\cal D} \delta h_j ~
  G_j[\delta h] ~ \widetilde{P} [\delta h,t].
\label{stochastic13}
\end{eqnarray}
The last step is to introduce the Fourier transform of
$p(\delta h_j,t)$ which can be reckoned as a generating function
for all moments of the distribution. Specifically on defining
\begin{eqnarray}
\widehat{p_j}(q,t) &=& \int_{- \infty}^{+\infty} ~d \delta h_j
~e^{\mathrm{i}q \delta h_j} p(\delta h_j,t),
\label{stochastic14}
\end{eqnarray}
we find a simple equation for the average $\widehat{p}(q,t)$ over all
sites of $\widehat{p_j}(q,t)$ :
\begin{eqnarray}
\partial_t \widehat{p}(q,t)&=& -2 D_{\mathrm{eff}} q^2  \widehat{p}(q,t)-
\mathrm{i} q  \widehat{\pi}(q,t),
\label{stochastic15}
\end{eqnarray}
in which $\widehat{\pi}(q,t)$ is the Fourier transform of $\pi(\delta h_j,t)$
averaged over all sites.
One can then expand Eq. (\ref{stochastic15}) in powers of $q$
and obtain an infinite hierarchy (closure problem)
in the correlation functions. The first two non-trivial orders ($O(q^2)$
and $O(q^3)$) are

\begin{eqnarray}
\frac {d g_0^{(2)}(t)}{dt}&=& 4 \nu _{\mathrm{eff}}
\Bigl[ g_1^{(2)}(t)-g_0^{(2)}(t)\Bigl]  + 4 D_{\mathrm{eff}},
\label{stochastic16}
\end{eqnarray}
and
\begin{eqnarray}
\frac {d g_{00}^{(3)}(t)}{dt}&=&- 3 \nu _{\mathrm{eff}}
\Bigl[ g_{11}^{(3)}(t)+g_{01}^{(3)}(t)-2 g_{00}^{(3)}(t)  \Bigl]
{+} \frac{1}{2} \lambda _{\mathrm{eff}}
\Bigl[g_{001}^{(4)}(t)-g_{111}^{(4)}(t) \Bigl],
\label{stochastic17}
\end{eqnarray}
where we have defined the following higher order correlation functions
\begin{eqnarray}
g_{lm}^{(3)}(t) &=& \frac{1}{N}  \sum_{i=1}^{N} \left \langle
\delta h_i(t) ~ \delta h_{i+l}(t)~ \delta h_{i+m}(t) \right \rangle,
\label{stochastic18}
\end{eqnarray}
\begin{eqnarray}
g_{lmn}^{(4)}(t) &=& \frac{1}{N} \sum_{i=1}^{N} \left \langle
\delta h_i(t) ~ \delta h_{i+l}(t)~ \delta h_{i+m}(t)
~\delta h_{i+n}(t) \right \rangle.
\label{stochastic19}
\end{eqnarray}

It is worth mentioning that
$\lambda $  does not explicitly  appear  in equation (\ref{stochastic16}).
As one can explicitly check, this is a feature associated to the particular
discretization Eq. (\ref{interface9})  and it would have {\it not} been the
case had we used the standard discretization for $F_i^{\lambda}[h]$.
This is clearly in turn related to the fact that the steady state
probability distribution Eq. (\ref{interface7}) ££ is independent of
$\lambda $.
We also note that in the $1+1$ case we are considering,
the explicit steady state solution of Eq.(\ref{stochastic15}) is known,
and depends only on
a single parameter $\frac{D}{\nu}$. As a consequence,  the steady
version of Eq.(\ref{stochastic15}) cannot be   used here to identify
$\nu$ and $D$.  In the $2+1$ case   where such a peculiar feature is
not present,  the stationary solution  depends on   $\lambda$ as well and
parameters identification can exploit the steady analogue equation.

\section{Parameters identification}
\label{sec:identification}

Our aim is to implement an identification procedure which could be
exploited on real experiments. For this reason, we assume that the
experimental surface is  constituted by a finite number of sites $N$
with lattice spacing $a$ (corresponding to a size $L=N a$),
and it is measured during  a  finite  time $T_{obs}$
every sampling time $\Delta t$.  Again $\Delta t$
is  a priori different from the discretization time $\delta t$ when the data is
produced numerically (note that in real experiments $\delta t$
is not even  defined!).
We  shall test the robustness and efficiency of the scheme with respect
to the size $L$ and  sampling time
$\Delta t$.

Identification methods are often based on minimizing a cost function defined
through dynamical constraints. This is clearly the case of the least square
method as explained in section~\ref{sec:least_squares}. Here we  derive
dynamical  constraints using Eqns. (\ref{stochastic5}) and
(\ref{stochastic16})
which contain informations of the original Langevin equation including mean
values and  fluctuations around  mean values. The present
identification is thus based    on  {\it deterministic}  equations. This
constitutes a  crucial difference with respect to the previous reconstruction
method \cite{Lam93}  directly based  on { \it stochastic} equations.
Another important
feature is that the observed quantities we use in our reconstruction scheme
are   dealing with averaged site values. Hence the fluctuations of all these
terms, which derive from stochastic quantities,  are {\it reduced}
typically by a factor $1/\sqrt{N}$, and self-averaging is expected to be more
effective.

Let us  derive the constraints we use. First, the total observation time
$T_{obs}$  is   divided into $q$ equal slices
$\bigl[T_1,T_2\bigr]$,$\ldots$,$\bigl[T_{q},T_{q+1}\bigr]$ with
$\Delta T = T_{j+1}-T_j$. Let us
integrate  (\ref{stochastic5})  and (\ref{stochastic16}) on each slice
$\bigl[T_{j},T_{j+1}\bigr]$:

\begin{eqnarray}
\frac {\Delta g^{(1)}}{T_{j+1}-T_{j} } &=& c +
\frac{\lambda_{\mathrm{eff}}}{6}  \frac {1}{T_{j+1}-T_{j}}
\int_{T_{j}}^{T_{j+1}}~ dt~ \Bigl[ 2 g_0^{(2)} (t)+
g_1^{(2)} (t) \Bigr],
\label{identification1}
\end{eqnarray}

\begin{eqnarray}
\frac {\Delta g_0^{(2)}}{T_{j+1}-T_{j} } &=&
4 \nu_{\mathrm{eff}}  \frac {1}{T_{j+1} -T_{j}}
\int_{T_{j}}^{T_{j+1}}~ dt~ \Bigl[ g_1^{(2)} (t)-
g_0^{(2)} (t) \Bigr] + 4 D_{\mathrm{eff}}.
\label{identification2}
\end{eqnarray}

If  functions and  integrals  in Eqns.
(\ref{identification1}) and (\ref{identification2})
are  computed using  experimental data, these discrete  equations
provide  $2q$ relations between the parameters to identify. From these
constraints, two cost functions are built  in a way already
described in Sec.\ref{sec:least_squares} with  $p=2$. The corresponding $2
\times 2$ equations then yield $c$ and $\lambda$ from one cost function and
$\nu$ and $D$ from the other.

We now explain how  functions and  integrals
in Eqns. (\ref{identification1}) and (\ref{identification2}) are obtained
experimentally. Starting with  a {\it same} initial surface   e.g. a flat
surface, we will  grow the   surface   ${\cal R}$ times. Because of the
stochastic nature of the phenomenon, this produces ${\cal R}$ different
observations or realizations of the same process .
Such a procedure, which can be performed very easily  in real
experiments,  allows the computation, at sampling times  $t=t_k=k \Delta t$
($k=1,\ldots,M+1$), of  $[g^{(1)}]_{\mathrm{exp}}$,
$[g_0^{(2)}]_{\mathrm{exp}}$
and $[g_1^{(2)}]_{\mathrm{exp}}$. Indeed, these quantities  are  the averages
over  ${\cal R}$ different realizations of  the spatial average height and
correlations of the first neighbors. The number ${\cal R}$ of realizations
need not be large: if  the total number $N $ of sites is sufficiently large,
the experimental values  are   rather close to the corresponding theoretical
predictions $g^{(1)}(t)$, $g_0^{(2)}(t)$  and $g_1^{(2)}(t)$. From these
functions sampled every $t=t_k=k \Delta t$,  integrals in Eqns.
(\ref{identification1}) and (\ref{identification2}) can be efficiently
evaluated  for small sampling time $\Delta t$.
In this case, the smooth functions $[g^{(1)}]_{\mathrm{exp}}$,
$[g_0^{(2)}]_{\mathrm{exp}}$  $[g_1^{(2)}]_{\mathrm{exp}}$
can  be approximated  on the whole time interval
$\bigl[T_1,T_{q+1}\bigr]$ by a standard curve fitting algorithm which
gives as a by product  the  time integrals.
This method   does impose   a constraint on
the sampling time $\Delta t$. However, this constraint  is substantially
weaker with respect to that imposed by the LS method
as we will show below. This
is a considerable advantage of our new procedure.

Two remarks are here in order.
Firstly one expects the result to be independent
on the  number of slices $q$  provided that $q$ satisfies the
following two constraints. On the one hand $q$ should be greater than $2$
(since two parameters are identified per
cost functions) and on the other hand it should   be less than
$ M= \frac {T_{obs}}{ \Delta t }$  so that  $ \Delta T $ cannot be less
than the sampling time  $\Delta t$.  Secondly the identification of
$\nu$ and $D$ could be achieved by using Eq.(\ref{stochastic15}) rather
than Eq. (\ref{stochastic16}). We shall see that in our approach the two
equations yield virtually identical results.

\section{Results}
\label{sec:results}

In order to test the potentiality of the different identification methods, we
produce experimental  data  by simulating   Eq.(\ref{interface5}) with a
standard Euler time integration algorithm with time step $\delta t=0.01$,
lattice spacing $a=1$, and
parameters $\nu=D=1$ and $\lambda=3$. These are the same values used in
Ref.\cite{Lam98_1}. The  time step is expected to be sufficiently small for not
causing  instability problems and the non-linear term $\lambda$ is big
enough to be well inside the KPZ phase. We find interesting to
repeat each calculation few times (typically 5) to give an estimate
of the error bars to be associated to each parameter value (this
was missing in previous works).

\subsection{LS Method}
\label{subsec:LS}

\begin{figure}[tbp]
\centerline{
\epsfxsize=3.0truein
\epsfysize=3.0truein
\epsffile{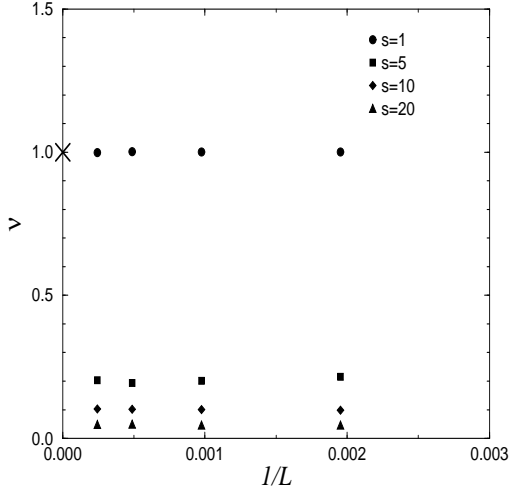}
}
\caption{The coupling parameter $\nu$ for increasing   lattice
sizes $L=512$, $1024$, $2048$, $4096$, in the  original steady-state LS method.
All quantities are in dimensionless form.
Error bars are of  order of  the symbol  sizes and
are consequently  not displayed. Different curves refer to increasing values of
the ratio  $s=\Delta t / \delta t$. The cross (X) indicates the exact value of
the  parameter $\nu =1$.}
\label{fig1}
\end{figure}

Let us  compute the parameters using the original LS method with the
spatial and temporal discretization Eqns. (\ref{interface5}) and
(\ref{interface6}).   We exploit  the
same trick used in Ref.\cite{Lam98_1} in which a KPZ surface of size $2L$ is
obtained by a magnification of a fully relaxed  surface of size $L$ where
the height are rescaled by a factor $2^{\alpha}$, ($\alpha=0.5$) and
linearly interpolated. The obtained surface is then relaxed to stationarity
before a successive magnification is attempted. However, unlike
Ref.\cite{Lam98_1} where a single surface of size $L=32768$ was computed,
we consider $L=512,1024,2048,4096$ and
linearly extrapolate the results to the limit  $L \to \infty$.
The calculation is repeated for increasing values of $s=\Delta t / \delta t$
in order to display the crucial weakness of the method as explained before.
Fig. \ref{fig1} depicts the results for
the parameter $\nu$ at finite $L$. Similar trends are present for
 $\lambda$ and $D$. The extrapolated values at $L \to \infty$ are reported
in Table
\ref{table1}. The gradual decrease in the precision of the reconstructed
parameters is apparent and it shows the loss of accuracy of the LS method
as $\Delta t$ increases as previously advertised.

We also considered the LS when the reconstructed quantities are computed
at time intervals $\Delta T$ which are multiple of the sampling time
$\Delta t$. In fact this test was also carried out by the authors of Ref.
\cite{Lam98_1} (in their notations $\tau = \Delta T$ and
$\Delta t=\delta t$) and
it will constitute a further source of comparison with our alternative
stochastic method (see below). Even in this case there is
a decrease in the performance of the procedure as the ratio
$r=\Delta T / \Delta t$ increases, consistently with the results
of Ref. \cite{Lam98_1}. The corresponding extrapolated values
are reported in Table \ref{table2}.

\subsection{Stochastic Approach}
\label{subsec:SA}

\begin{figure}[tbp]
\centerline{
\epsfxsize=3.0truein
\epsfysize=3.0truein
\epsffile{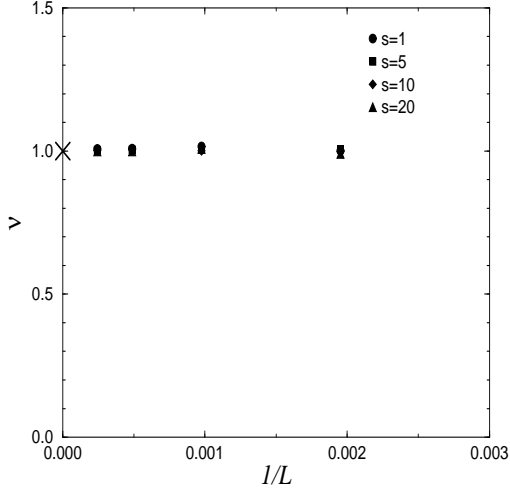}
}
\caption{The coupling parameter $\nu$ for increasing   lattice
sizes $L=512$, $1024$, $2048$, $4096$, as obtained
from our reconstruction method. Error bars are of  order of  the symbol  sizes
and   are consequently  not displayed. Different curves refer to increasing
values of the ratio $s= \Delta t / \delta t$. The
cross (X) indicates the exact value of the parameter  $\nu =1$.}
\label{fig2}
\end{figure}

For a more convenient comparison with the LS method, we use the same sizes and
statistics (five different configurations for each size). Our calculations
are carried out in the {\it transient} rather than in
the {\it steady} state and are therefore much less time consuming.  Again the
results are obtained for  $L=512,1024,2048,4096$ and  linearly extrapolated to
$L \to \infty$. For a comparison with the previous calculation, the outcomes
of the parameter $\nu$ at different size
$L$ are plotted in Fig.\ref{fig2}
for increasing values of the ratio $s = \Delta t / \delta t$,
and the corresponding extrapolated values are reported
in Table \ref{table3}. One can see
that the parameter values are rather  insensitive to the changing the ratio
$s= \Delta t / \delta t$ as expected.
Next we check the performance of our method with respect to the
increasing of the ratio $r=\Delta T / \Delta t$. This is reported
in Table \ref{table4}.
As expected, our method outperforms the LS one in all situations.

Since the LS method could in principle be carried out in
transient rather than
in steady state conditions, one might wonder how it would perform
in this case.

To this aim we recompute the parameters using the LS method under these
conditions and find that the predicted values are far off
with respect to
the exact ones. For instance for $L=4096$ a typical run yields $\nu \sim 0.36$,
$\lambda \sim 0.68$ and $D \sim 0.005$ to be compared with a typical
result obtained with our method $\nu \sim 0.99$,
$\lambda \sim 2.98$ and $D \sim 1.01$.

As a final cross-check of our method, we recompute the parameters in the same
situation as before but using Eq. (\ref{stochastic15}) rather than Eq.
(\ref{stochastic16}) to extract $\nu$ and $D$ and find nearly identical values.

\subsection{Coarse-graining and KPZ real discretization.}
\label{subsection:CG}

The application of this machinery to experimental surfaces  assumes that
the system is described by a KPZ-like dynamics.
In this case, besides
being able to address the issue of whether or not they belong to the KPZ
universality class, one would be able to provide a numerical estimates of the
coupling parameters which are usually overlooked in studies focussing only
on the
universality class.

Following Lam and Sander \cite{Lam93}, we produce an interface
based on the KPZ discretized model Eq. (\ref{interface5}) which
is then smoothed by introducing the (discrete) Fourier transform of the heights
\begin{eqnarray}
\widehat{h}_{q_{n}}(t) &=& a \sum_{i=1}^{N} e^{\mathrm{-i}q_n x_i}
h_i(t)
\label{results1}
\end{eqnarray}
A coarse-graining surface at level $a_s = b a$ can then be achieved
by simplying
setting to zero all wavelength components $\widehat{h}_{q_{n}}(t)$ with
$q \ge N_s=L/a_s$ and transforming back to real space.
The obtained smoothed surface
is then assumed to be governed by a KPZ equation (renormalizability
property). This new KPZ dynamics can then reconstructed along lines
similar to those described above before a further time step is carried
out on the original surface.

With $b=2$, $\Delta T/ \Delta t=1$ and sizes up to $L=8192$
averaged over 5 configurations as before,
we find $\nu=1.09 \pm 0.04$, $\lambda =3.27 \pm 0.05$, and $D=0.88 \pm 0.03$.
Higher values of $b$ result in poorer and poorer agreement with the expected
values even with higher lattice sizes.
The same feature is also present in the original LS procedure
as we explicitly checked.
In fact this is a general deficiency
of the real space discretization as explained in Appendix \ref{sec:appendixb}:
the finite size difference has lost some renormalizability property of
the original KPZ continuum equation.

\section{Conclusions}
\label{sec:conclusions}
In this paper, we discuss a method for extracting the coupling parameters from
a non-linear Langevin equation starting from  experimental surfaces
representing successive snapshots of the system.
We apply this scheme to the KPZ equation
in $1+1$ dimensions (although it could be extended to any dimensions)
and compare it with the previous approach of Ref. \cite{Lam93}, finding the
following differences.  First of all it does not require large sizes
and it is well suited for a transient state.
This is expected to be a considerable advantage, notably in
numerical work, since the  typical time required to reach a steady state
increases as $L^z$ where $z$ is the  dynamical exponent ($3/2$ in the $1+1$ KPZ
case).  We have explicitly shown how the LS method which works rather well in
the aforementioned conditions, fails to provide sensible answers otherwise.
Most importantly however is the fact that our approach is stable
under the changing of the sampling time, unlike the LS method which is not.
We stress the importance of this feature since in typical experimental
situations, the sampling time is an externally tuned parameters
which has nothing to do with the
evolution time of the system. We have discussed the reasons why this is
so and provide an intuitive  heuristic argument showing why the LS scheme
is {\it not} expected to work under these more realistic conditions.
Finally we implemented a coarse-graining procedure in order to be
able to apply our method to experimentally generated profiles. We showed
that the agreement with the expected values is much poorer in the
present case and we further argued that
{\it any} real space based approach is doomed to run into this problem, the
reason being
that they have not a correct renormalization behavior under coarse-graining as
explained in Appendix \ref{sec:appendixb}. We have recently devised an
alternative approach based on a Fourier-based scheme which avoids discretization
problems. The results of this will be the subject of a forthcoming publication.

\acknowledgments
This work was supported by a joint CNR-CNRS exchange program number 5274.
One of us
(AG) acknowledges financial support by MURST and INFM.

\appendix
\section{A simple solvable example}
\label{sec:appendixa}

This   appendix shows, on a simple and solvable example which is a
zero-dimensional analogue of Eq.(\ref{interface1}), that
the method of least squares can be hampered by the presence of dynamical noise.
Let us assume that the scalar variable $X(t)$ is governed by the following
Langevin equation:
\begin{eqnarray}
\frac{d X  }{dt}&=&  B (X )+ \mu G (X ) +  \eta(t),
\label{appendixa1}
\end{eqnarray}
where  $B $ and $G  $ are  prescribed functions, $\eta(t)$ is  an uncorrelated
white noise
\begin{eqnarray}
\left \langle \eta (t) \eta (t') \right \rangle &=&
2 D \delta(t-t').
\label{appendixa2}
\end{eqnarray}

For simplicity, we assume  that: (a) measurement noise is negligible, (b)
the observed time serie $X^{\mathrm{obs}}(t_k)$ with $t_k = k \Delta t$
($k=1,\ldots,M+1$)  has been produced by the dynamical system
(\ref{appendixa1}) with  the  value $\mu =0$.  We ask whether the least square
method is capable of identifying  the  correct coupling parameter $\mu =0$.

From the one hand, the least square method first assumes that the data is
produced by the deterministic counterpart of Eq.(\ref{appendixa1})  with an
unkown parameter $\mu$. In discrete times, this  yields a "predicted" value
given by
\begin{eqnarray}
X^{\mathrm{pred}} (t_{k+1}) & = & X^{\mathrm{obs}}(t_k) +
\Delta t \Bigl[  B (X^{\mathrm{obs}}(t_k))  + \mu
G(X^{obs}(t_k)) \Bigr].
\label{appendixa3}
\end{eqnarray}
On the other hand, the "observed" value is given,  if the sampling time is small
enough,   by the discrete time counterpart of Eq.(\ref{appendixa1}) with
$\mu=0$ :
\begin{eqnarray} X^{\mathrm{obs}}(t_{k+1}) & =&
X^{\mathrm{obs}}(t_k) + \Delta t     ~ B (X^{obs}(t_k) )
+ r(t_k) \sqrt {2 D \Delta t},
\label{appendixa4}
\end{eqnarray}
where $r(t_k)$ is a gaussian random generator of unit variance.

Using both experimental observations $X^{\mathrm{obs}}(t_k)$ and
reconstructed quantities $X^{\mathrm{pred}}(t_k) $,  a cost function can be
constructed:
\begin{eqnarray}
{\cal J}&=&\frac{1}{M} \sum_{k = 1}^{M}   \Bigl[ X^{\mathrm{obs}}(t_{k+1}) -
X^{\mathrm{pred}} (t_{k+1}) \Bigr]^{2}.
\label{appendixa5}
\end{eqnarray}
Using Eqns. (\ref{appendixa3}) and (\ref{appendixa4}), the cost function
is readly rewritten as
\begin{eqnarray}
{\cal J} &=& \frac{1}{M} \sum_{k = 1}^{M}   \Bigl[ \Delta t  \mu
G(X^{\mathrm{obs}}(t_k) ) -  r(t_k) \sqrt { 2 D \Delta t}   \Bigr]^{2}.
\label{appendixa6}
\end{eqnarray}

The  minimum value of ${\cal J}$  then satisfies the extremality condition

\begin{eqnarray}
\frac{\partial {\cal J} }{\partial \mu  }|_{\mu^{*}} &=&0,
\label{appendixa7}
\end{eqnarray}
which provides the value
\begin{eqnarray}
\mu^{*} &=& \sqrt{\frac{2D}{\Delta t}}  \frac{ \frac{1}{M}
\sum_{k=1}^{M} r(t_k)
G(X^{\mathrm{obs}}(t_k))}{ \frac{1}{M} \sum_{k = 1}^{M}
G^2(X^{\mathrm{obs}}(t_k) )}.
\label{appendixa8}
\end{eqnarray}

Let us  estimate the two  sums appearing in Eq.(\ref{appendixa8}).
Because of self-averaging, the denominator can be clearly rewritten as
\begin{eqnarray}
\frac{ 1} {M} \sum_{k = 1}^{M} G^2(X^{\mathrm{obs}}(t_k))  \approx \left
\langle
G^2 \right \rangle ,
\label{appendixa9}
\end{eqnarray}
the average being over the noise $\eta$.
The second sum in can be separated into two contributions
\begin{eqnarray}
\frac{ 1}{M} \sum_{k = 1}^{M} \Bigl[ r(t_k) G(X^{\mathrm{obs}}(t_k)]
&=& \frac{ 1}{M} \sum_{k=1}^{M}
r(t_k) \Bigl[ G(X^{obs}(t_k)- \left \langle G \right \rangle \Bigr] +
\frac{ \left \langle G \right \rangle}{M} \sum_{k = 1}^{M},
r(t_k)
\label{appendixa10}
\end{eqnarray}
where using the Central Limit Theorem (CLT) \cite{Gardiner90}
we have that
\begin{eqnarray}
 \frac{\left \langle G \right \rangle}{M} \sum_{k = 1}^{M} r(t_k) &\approx &
 \frac{ \left \langle G \right \rangle }{\sqrt M} r_1 ,
\label{appendixa11}
\end{eqnarray}
where $r_1$ is a random variable of unit variance.
In Eq.(\ref{appendixa10}), each term of the first sum of the r.h.s. is a
random variable which is  a
product of
two independent random variables of zero mean and variance respectively
equal to $1$ and
${\sigma}  = \left \langle G^2 \right \rangle -\left \langle G \right
\rangle^2$.
A similar argument based on the CLT applies to
the evaluation of the first  term of the r.h.s. since
it  is again a sum of  $M$ random variables with zero average which implies
that:
\begin{eqnarray}
\frac{ 1}{M} \sum_{k = 1}^{M}
r(t_k) \Bigl[ G(X^{\mathrm{obs}}(t_k)-\left \langle G \right \rangle \Bigr]
& \approx & \frac{\sqrt{\sigma}}{ \sqrt M} r_2 ,
\label{appendixa12}
\end{eqnarray}
where $r_2$ is a random variable of unit variance.
Instead of the true value one then finds
\begin{eqnarray}
\mu^{*} &\approx&  \frac{  \sqrt {2 D} }{ \sqrt { M \Delta t  } }
 \frac { Max \Bigl( \sqrt{\sigma}, \left \langle G \right \rangle \Bigr)}{
\left \langle
G^2 \right \rangle }.
\label{appendixa13}
\end{eqnarray}
With non-vanishing amplitudes $D$, the noise   hence causes errors  for
small $\Delta t $
unless  large statistics are considered.

\section{Non-renormalizability of real space discretizations}
\label{sec:appendixb}

Let us consider a one-dimensional surface defined on a lattice of length
$L=N~a$ ($a$ being the lattice spacing and $N$ being the total number of
sites) which is identified by $h_1,\ldots,h_N$ at positions
$x_1=a,\ldots,x_N=Na$ and having periodic boundary
conditions.
If we assume a KPZ dynamics, then the equation of motion is
given by Eq.(\ref{interface5}) and its stationary state by Eq.
(\ref{interface7}). Let us introduce the (discrete) Fourier
transform $\widehat{h}_q$ so  that
\begin{eqnarray}
h_i(t) &=& \frac{1}{L} \sum_{n=-N/2}^{N/2} e^{\mathrm{i}q_n x_i}
\widehat{h}_{q_{n}}(t) ,
\label{appendixb1}
\end{eqnarray}
and conversely
\begin{eqnarray}
\widehat{h}_{q_{n}}(t) &=& a \sum_{i=1}^{N} e^{\mathrm{-i}q_n x_i} .
h_i(t)
\label{appendixb2}
\end{eqnarray}
By using the relation
\begin{eqnarray}
\sum_{i=1}^N  e^{\mathrm{i}\Bigl(q_n-q_m\Bigr) x_i} &=& N \delta_{n,-m} ,
\label{appendixb3}
\end{eqnarray}
it is easy to show that the correct corresponding of Eq.(\ref{interface7})
for the stationary distribution is
\begin{eqnarray}
\widehat{P_a}[\widehat{h}] &=& {\cal N}^{-1} \exp
\Biggl[ - \frac{1}{2}\frac{\nu}{DL} \sum_{n=-N/2}^{N/2} q_n^2
|\widehat{h}_{q_{n}}|^2 \Biggr] .
\label{appendixb4}
\end{eqnarray}
In Eq.(\ref{appendixb4}) the continuum limit $a \to 0$ is simply achieved
by letting
$N \to \infty$.

We now recall that the proper variables to be used in this
context are the $\delta h_i =h_{i}-h_{i+1}$ whose Fourier
transform $\delta \widehat{h}_{q_{n}}$ are related to the Fourier transform
$\widehat{h}_{q_{n}}$ of the heights by
\begin{eqnarray}
\delta \widehat{h}_{q_{n}} &=&\widehat{h}_{q_{n}}
\Bigl(1-e^{\mathrm{i}q_n a} \Bigr) .
\label{appendix5}
\end{eqnarray}
We can exploit the periodic boundary conditions to extend the $N-1$ sites
to $N$ (bearing however in mind that only $N-1$ are independent)
and use the the fact that
\begin{eqnarray}
\frac{1}{a} \sum_{i=1}^N \delta h_i^2 &=&
\frac{1}{La^2}
\sum_{n=-N/2}^{N/2} |\delta \widehat{h}_{q_{n}}|^2 ,
\label{appendix6}
\end{eqnarray}
to rewrite the probability (\ref{appendixb4}) in the alternative form
\begin{eqnarray}
\widehat{P_a}[\widehat{h}] &=& {\cal N}^{-1} \exp
\Biggl[ - \frac{1}{2}\frac{\nu}{D L a^2} \sum_{n=-N/2}^{N/2}
|\widehat{h}_{q_{n}} \Bigl( 1-e^{\mathrm{i} q_n a} \Bigr)|^2 \Biggr] .
\label{appendixb7}
\end{eqnarray}
In the continuum limit $a \to 0$  one can explicitly check that
Eq.(\ref{appendixb7}) is equivalent to Eq.(\ref{appendixb4}) by expanding
the term $e^{\mathrm{i} q_n a}$ in powers of $a$ and keeping the
lowest non-vanishing order $O(a)$. Nevertheless Eq.(\ref{appendixb7}) leads
to problems when one tries to coarse grain the surface. Indeed suppose
we now perform a smoothing of the lattice of a factor $b$ to obtain
a new lattice constant $a_s = b a$ and a decreased number of sites
$N_s=N/b$. This amounts to set to zero all modes from $N_s$ to $N$
and thus the {\it correct} corresponding stationary distribution is
according to (\ref{appendixb7})
\begin{eqnarray}
\widehat{P_{a_{s}}}[\widehat{h}] &=& {\cal N}_{\mathrm{s}}^{-1} \exp
\Biggl[ - \frac{1}{2}\frac{\nu}{D L a^2} \sum_{n=-N_s/2}^{N_s/2}
|\widehat{h}_{q_{n}} \Bigl( 1-e^{\mathrm{i} q_n a} \Bigr)|^2 \Biggr] .
\label{appendixb8}
\end{eqnarray}

On the other hand, had we started from the ``smoothed'' lattice
and construct the differences $\delta h_i^{\mathrm{s}}=
h_{i}^{\mathrm{s}}-h_{i+b}^{\mathrm{s}}$ of the coarse-grained heights,
we would have found
\begin{eqnarray}
\widehat{P_{a_{s}}}[\widehat{h}] &=& {\cal N}_{\mathrm{s}}^{-1} \exp
\Biggl[ - \frac{1}{2}\frac{\nu}{D L a_s^2} \sum_{n=-N_s/2}^{N_s/2}
|\widehat{h}_{q_{n}} \Bigl( 1-e^{\mathrm{i} q_n b a} \Bigr)|^2 \Biggr] ,
\label{appendixb9}
\end{eqnarray}
which differs from the previous one.

It is clear that this is a general problem of all discretizations
in real space.
In the limit $a \to 0$ in fact,
one recovers from (\ref{appendixb9}) the correct expression (\ref{appendixb4}).

\begin{table}
\caption{Extrapolated values of coupling parameters $\nu(\infty)$,
$\lambda (\infty)$ and $D(\infty)$ as a function
of $s=\Delta t / \delta t$ (see text) as computed from the
original LS method (at steady state).}
\begin{tabular}{lccc}
\multicolumn{1}{l}{$\Delta t / \delta t$}&
\multicolumn{1}{c}{$\nu(\infty)$}&
\multicolumn{1}{c}{$\lambda(\infty)$}&
\multicolumn{1}{c}{$D(\infty)$}\\
\hline
$1$  & $0.999 \pm 0.003$ & $2.980 \pm 0.005$ &$1.000 \pm 0.001$  \\
$5$  & $0.194 \pm 0.004$ & $0.595 \pm 0.015$ &$0.200 \pm 0.001$  \\
$10$ & $0.103 \pm 0.004$ & $0.319 \pm 0.008$ &$0.100 \pm 0.001$  \\
$20$ & $0.049 \pm 0.002$ & $0.139 \pm 0.004$ &$0.050 \pm 0.001$  \\
\end{tabular}
\label{table1}
\end{table}

\begin{table}
\caption{Extrapolated values of coupling parameters $\nu(\infty)$,
$\lambda (\infty)$ and $D(\infty)$ as a function
of $r=\Delta T / \Delta t$ (see text) as computed from the
original LS method(at steady state).}
\begin{tabular}{lccc}
\multicolumn{1}{l}{$\Delta T / \Delta t$}&
\multicolumn{1}{c}{$\nu(\infty)$}&
\multicolumn{1}{c}{$\lambda(\infty)$}&
\multicolumn{1}{c}{$D(\infty)$}\\
\hline
$1$  & $0.999 \pm 0.003$ & $2.980 \pm 0.005$ &$1.000 \pm 0.001$  \\
$5$  & $0.996 \pm 0.002$ & $2.817 \pm 0.004$ &$0.928 \pm 0.001$  \\
$10$ & $0.917 \pm 0.002$ & $2.627 \pm 0.006$ &$0.853 \pm 0.001$  \\
$20$ & $0.855 \pm 0.002$ & $2.308 \pm 0.006$ &$0.757 \pm 0.001$  \\
$50$ & $0.650 \pm 0.001$ & $1.558 \pm 0.006$ &$0.653 \pm 0.002$  \\
\end{tabular}
\label{table2}
\end{table}

\begin{table}
\caption{Extrapolated values of coupling parameters $\nu(\infty)$,
$\lambda (\infty)$ and $D(\infty)$ as a function
of $s= \Delta t / \delta t$ as computed from our stochastic approach
in the transient state.
}
\begin{tabular}{lccc}
\multicolumn{1}{l}{$\Delta t / \delta t$}&
\multicolumn{1}{c}{$\nu(\infty)$}&
\multicolumn{1}{c}{$\lambda(\infty)$}&
\multicolumn{1}{c}{$D(\infty)$}\\
\hline
$1$ & $1.009 \pm 0.002$ & $3.047 \pm 0.016$ &$1.026 \pm 0.001$  \\
$5$ & $1.008 \pm 0.007$ & $3.015 \pm 0.006$ &$1.003 \pm 0.007$  \\
$10$ &$1.035 \pm 0.011$ & $2.993 \pm 0.010$ &$1.018 \pm 0.011$  \\
$20$ &$0.997 \pm 0.020$ & $3.001 \pm 0.005$ &$0.978 \pm 0.010$  \\
\end{tabular}
\label{table3}
\end{table}

\begin{table}
\caption{Extrapolated values of coupling parameters $\nu(\infty)$,
$\lambda (\infty)$ and $D(\infty)$ as a function
of $r= \Delta T / \Delta t$ as computed from our stochastic approach
in the transient state.
}
\begin{tabular}{lccc}
\multicolumn{1}{l}{$\Delta T / \Delta t$}&
\multicolumn{1}{c}{$\nu(\infty)$}&
\multicolumn{1}{c}{$\lambda(\infty)$}&
\multicolumn{1}{c}{$D(\infty)$}\\
\hline
$1$  & $1.009 \pm 0.002$ & $3.047 \pm 0.016$ &$1.026 \pm 0.001$  \\
$5$  & $1.003 \pm 0.003$ & $3.005 \pm 0.010$ &$1.057 \pm 0.003$  \\
$10$ & $1.003 \pm 0.001$ & $3.030 \pm 0.007$ &$1.023 \pm 0.001$  \\
$20$ & $0.998 \pm 0.002$ & $3.014 \pm 0.009$ &$1.016 \pm 0.001$  \\
$50$ & $0.991 \pm 0.003$ & $3.017 \pm 0.006$ &$1.010 \pm 0.003$  \\
\end{tabular}
\label{table4}
\end{table}



\begin{references}

\bibitem{Weigend94} A.S. Weigend and N.A. Gershenfeld,   {\it Time Series
Prediction},
Addison Wesley, (1994).

\bibitem{Abarbanel93} H.D. Abarbanel, R. Brown, J. J. Sidorowitch and
LS Tsimring, Rev. Mod. Phys. {\bf 65}, 1331 (1993).

\bibitem{Kostelich93} E. J. Kostelich and T. Schreiber, Phys. Rev. E
{\bf 48}, 1752 (1993).

\bibitem{Gardiner90} See e.g. C. W. Gardiner {\it Handbook of Stochastic
Methods}, 2nd edition (Springer Verlag 1990); N. G. van Kampen
{\it Stochastic processes in Physics and Chemistry} (North-Holland 1992).

\bibitem{Battiston99} L. Battiston and M. Rossi, Int. Journal of Chaos
and Applications, (to appear)

\bibitem{Kardar86} M. Kardar, G. Parisi and Y. C. Zhang, Phys. Rev. Lett.
{\bf 56}, 889 (1986).

\bibitem{Krug97} For recent reviews see e.g.
J. Krug, Adv. Phys. {\bf 46}, 139 (1997);
A.L. Barabasi, H. E. Stanley, {\it Fractal Concepts in
Surface Growth} (Cambridge University Press, Cambridge 1995);
T. Halpin-Haley and Y. C. Zhang, Phys. Rep. {\bf 254}, 215 (1995);
M. Marsili, A. Maritan, F. Toigo and J. R. Banavar, Rev. Mod. Phys.
{\bf 68}, 963 (1996); P. Meakin, Phys. Rep. {\bf 235}, 131 (1993);

\bibitem{Edward82} S. F. Edward and D. R. Wilkinson, Proc. R. Soc. London
A {\bf 381}, 17 (1982).

\bibitem{Huse85} D. A. Huse and C. Henley, Phys. Rev. Lett. {\bf 54},
2708 (1985).

\bibitem{Forster77} D. Forster, D. R. Nelson and M. J. Stephen,
Phys. Rev. A {\bf 16}, 732 (1977).

\bibitem{Yakhot81} V. Yakhot, Phys. Rev. A {\bf 24}, 642 (1981).

\bibitem{Bogosian99} B. Bogosian, C. C. Chow and T. Hwa, cond-mat/9911069.

\bibitem{Lam93}C. Lam and L.M. Sander, Phys.Rev. Lett. {\bf 71 },
561, (1993).

\bibitem{Lam98_1} C. Lam and F.G. Shin, Phys.Rev. E {\bf 58 },5592 (1998).

\bibitem{Risken89} H. Risken,
{\it The Fokker-Planck Equation}, Springer-Verlag,
Berlin, (1989).

\bibitem{Beccaria94} M. Beccaria and G. Curci, Phys. Rev. E {\bf 50},
4560 (1994)

\bibitem{Newman97} T. J. Newman and M. R. Swift, Phys. Rev. Lett.
{\bf 79}, 2261 (1997)

\bibitem{Newman96} T. J. Newman and A. J. Bray, J. Phys. A
{\bf 29}, 7917 (1996).

\bibitem{Lam98_2} C. Lam and F.G. Shin, Phys.Rev. E {\bf 57 },6506 (1998).

\bibitem{Mannella89}
R.Mannella, {\it Computer experiments in non-linear
stochastic physics}, in: {\it Noise in nonlinear dynamical systems}, vol.
3, ed. by
F. Moss, P.V.E. McClintock, Cambridge University Press, Cambridge, (1989).

\bibitem{note1} A correct discretization was also previously
obtained by A. Maritan (private communication).
\bibitem{Goldenfeld93} See e.g. N. Goldenfeld {\it Lectures on
Phase Transitions and the Renormalization Group} (Addison-Wesley
1993) pg. 226.
\end{references}
\end{document}